\documentclass{ECOC-authorr}
\usepackage{etex}
\usepackage{amsmath,amssymb}
\usepackage{tikz}
\usetikzlibrary{decorations}
\usetikzlibrary{calc} 
\tikzset{>=latex}
\usetikzlibrary{arrows,shapes}
\usetikzlibrary{arrows}
\usetikzlibrary{plotmarks}
\usepackage{ctable}
\usepackage{makecell}
\usepackage{caption}

\newcommand{\nc}{n}
\newcommand{\kc}{k}
\newcommand{\BB}{\mathsf{B}}
\newcommand{\cc}{\boldsymbol{C}}

\newcommand{\lcc}{\boldsymbol{c}}
\newcommand{\lrr}{\boldsymbol{r}}
\newcommand{\w}{w}

\newcommand{\AS}{\textcolor{black}}
\begin{document}
	
	\title{ON PARAMETER OPTIMIZATION OF PRODUCT CODES FOR ITERATIVE BOUNDED DISTANCE DECODING WITH SCALED RELIABILITY}
	
	\author{Alireza Sheikh\ad{1,2*}, Alexandre Graell i Amat\ad{1}, Gianluigi Liva\ad{3}, and  Alex Alvarado\ad{2}} 
	
	\address{\add{1}{Department of Electrical Engineering, Chalmers University of Technology, Gothenburg, Sweden}
		\add{2}{Department of Electrical Engineering, Eindhoven University of Technology, Eindhoven, The Netherlands}
		\add{3}{Institute of Communications and Navigation of the German Aerospace Center (DLR), Munich, Germany}
		\email{asheikh@chalmers.se}}
	
	
	\begin{abstract}
		We use density evolution to optimize the parameters of binary product codes (PCs) decoded based on the recently introduced iterative bounded distance decoding with scaled reliability. We show that binary PCs with component codes of $3$-bit error correcting capability provide the best performance-complexity trade-off.
	\end{abstract}
	
	\maketitle
	\section{Introduction}
	\vspace{1ex}
	Product-like codes such as product codes (PCs) and staircase codes are considered extensively in the optical community due to their excellent performance. Soft decision decoding (SDD) of PCs provides high net coding gains at the cost of high decoding data flow and power consumption \cite{Pyn98,Fougstedt2019}. Due to the stringent power constraints of fiber-optic systems operating at $400$ Gbps and beyond  \cite{Agrell_2016}, hard decision decoding (HDD) based on bounded distance decoding (BDD) of the component codes is an appealing alternative. HDD yields good performance with significantly lower decoding data flow and power consumption \cite{Fougstedt2019}. 
	
	An interesting line of research is to reduce the gap between SDD and HDD while limiting the complexity increase with respect to HDD. In particular, concatenation schemes based on inner SDD and outer HDD have been proposed where the inner codes are designed to have a constrained complexity \cite{Ksch17,Ksch18}. An alternative approach is to assist the HDD with some level of soft information to improve its performance, while maintaining the decoding complexity and data flow similar to that of HDD. Recently, new soft-aided decoding algorithms have been proposed for product-like codes \cite{She18,She19Tcom,Yibitflip,She19}, enabling an attractive solution for future high-throughput, low-power fiber-optic systems. In particular, in \cite{She18,She19Tcom} a novel decoding algorithm, called iterative BDD with scaled reliability (iBDD-SR) which utilizes the channel reliabilities to diminish the number of miscorrections was proposed. 
	An implementation architecture based on $28$ nm technology in \cite{FougstedtiBDDSR2018} showed that iBDD-SR with PCs for an overhead (OH) of $21.9 \%$ achieves energy efficiency of $0.63$ pJ/bit for $1$ Tbps information throughput. This is less than half of the corresponding decoder energy efficiency of staircase codes decoded using conventional iterative BDD \cite{FougstedtiBDDSR2018,Fougstedt2019}.\\ 
	\indent In order to use PCs with iBDD-SR in fiber-optic systems, one needs to select the parameters of the underlying component code for a target OH. Component codes with different lengths and error correction capabilities can lead to the same OH. A naive approach for code optimization is to simulate the performance of all possible PCs for a target OH, which is computationally intensive. An alternative approach is to predict the performance of PC analytically, and select its parameters accordingly.
	
	
	In this paper, we optimize the parameters of binary PCs for various OHs, ranging from $6.25 \%$ to $33.33 \%$, using the density evolution (DE) analysis recently derived in \cite{She19Tcom}. In particular, we adapt the DE for iBDD-SR to the case where the component code of a PC is shortened, in order to achieve different OHs.  We show that PCs built based on component codes with error correction capability of $3$ bits provide the best performance-complexity trade-off for iBDD-SR and all OHs of interest. Furthermore, the performance of optimized PCs approaches that of the iterative miscorrection-free decoder. Using the Gaussian noise (GN) model, we show that the performance improvement of optimized PCs decoded with iBDD-SR over iterative BDD yields optical reach enhancement up to $5.8 \%$ of the original reach achieved with conventional iterative BDD for a wavelength-division multiplexing (WDM) system with quadrature phase shift keying (QPSK).
	\newcommand{\tablehighlight}{}
	\begin{table*}[t]
		\caption{Optimized parameters of PCs based on DE analysis for iBDD-SR and iBDD (within parenthesis). ``k'' stands for $10^{3}$.}
		\centering
		\setlength{\tabcolsep}{6.3pt} 
		\renewcommand{\arraystretch}{0.88} 
		\small
		\scalebox{0.8}{
			\begin{tabular}{c|c|c|c|c|c|c|c|c|c|c|c|c|c|c}
				\toprule
				\midrule
				\tablehighlight{OH (\%)} & \tablehighlight{$33.33$} &
				\tablehighlight{$25.00$} &
				\tablehighlight{$20.00$} & \tablehighlight{$16.67$} &
				\tablehighlight{$14.29$} & \tablehighlight{$12.50$} & \tablehighlight{$11.11$} & \tablehighlight{$10.00$} & \tablehighlight{$9.09$} & \tablehighlight{$8.33$} & \tablehighlight{$7.69$} & \tablehighlight{$7.14$} & \tablehighlight{$6.67$} & \tablehighlight{$6.25$} \\
				\hline
				\tablehighlight{$v$} & $8$ ($8$) & $8$ ($8$) & $9$ ($9$) & $9$ ($9$) & $9$ ($9$) & $9$ ($9$) & $9$ ($9$) & $10$ ($10$) & $10$ ($10$) & $10$ ($10$) & $10$ ($10$) & $10$ ($10$) & $10$ ($10$) & $10$ ($10$) \\
				\tablehighlight{$t$} & $3$ ($4$) & $3$ ($3$) & $3$ ($4$) & $3$ ($4$) & $3$ ($3$) & $3$ ($3$) & $3$ ($3$) & $3$ ($4$) & $3$ ($4$) & $3$ ($4$) & $3$ ($4$) & $3$ ($3$) & $3$ ($3$) & $3$ ($3$) \\		
				\tablehighlight{$s$} & $76$ ($16$) & $28$ ($28$) & $201$ ($98$) & $147$ ($26$) & $93$ ($93$) & $39$ ($39$) & $0$ ($0$) & $378$ ($163$)  & $318$ ($83$) & $258$ ($3$) & $198$ ($0$) & $138$ ($138$) & $78$ ($78$) & $18$ ($18$) \\
				\hline
				\tablehighlight{$n_\mathsf{c} \approx$} & \makecell{$32$k\\ $(57\text{k})$} & \makecell{$52$k\\ $(52\text{k})$} & \makecell{$96$k\\ $(170\text{k})$} & \makecell{$132$k\\ $(235\text{k})$} & \makecell{$175$k\\ $(175\text{k})$} & \makecell{$223$k\\ $(223\text{k})$} & \makecell{$261$k\\ $(261\text{k})$} & \makecell{$404$k\\ $(740\text{k})$}  & \makecell{$497$k\\ $(884\text{k})$} & \makecell{$585$k\\ $(1040\text{k})$} & \makecell{$681$k\\ $(1047\text{k})$} & \makecell{$783$k\\ $(783\text{k})$} & \makecell{$893$k\\ $(893\text{k})$} & \makecell{$1010$k\\ $(1010\text{k})$} \\		
				\midrule
				\bottomrule
			\end{tabular}
		}
		\vspace{-2ex}
		\label{tab:Table1}
	\end{table*}
	\vspace{0.5ex}

	\section{Preliminaries, PC decoding algorithms}
	Let $\mathcal{C}$ be a Bose-Chaudhuri-Hocquenghem (BCH) component code built over the Galois field (GF) of order $2^v$, GF($2^v$), with length $\nc=2^v-1-s$, dimension $\kc=2^v-vt-1-s$, error correction capability $t$, and shortening parameter $s$; in short $(v,t,s)$. A two-dimensional binary PC with length $n_\mathsf{c}=n^2$ and $\text{OH}~=~\nc^2/\kc^2-1$ is defined as the
	set of all $\nc\times\nc$ arrays such that each row and column  is a codeword of $\mathcal{C}$. Each codeword is represented by a binary matrix $\cc=[c_{i,j}]$.
	We consider the binary-input additive white Gaussian noise channel, where $L_{i,j}$ is the channel log-likelihood ratio (LLR) corresponding to $c_{i,j}$. The hard decision on the channel output is denoted by $r_{i,j}$ and obtained by
	mapping the sign of $L_{i,j}$ according to $- 1 \mapsto 1$ and $+ 1
	\mapsto 0$ using a mapping function $\BB(\cdot)$, i.e., $r_{i,j}=\BB(L_{i,j})$. 
	
	PCs are usually decoded using BDD of
	the component codes.
	Let $\lcc=(c_1,\ldots,c_{n})$ be the transmitted codeword of length $n$ of a component code and $\lrr=(r_1,\ldots,r_{n})$ the received hard-detected bits. 
	BDD corrects all
	error patterns with Hamming weight up to $t$. If the
	number of errors in $\lrr$ is larger than $t$ and there exists
	another codeword $\tilde{\lcc} \in \mathcal{C}$ with
	Hamming distance to $\lrr$ less than $t$, BDD outputs $\tilde{\lcc}$ and a miscorrection occurs.  Otherwise, BDD fails and the decoder outputs $\lrr$. 
	We denote the iterative decoding of PCs based on BDD of row and column codes as iBDD.
	We use the term ``ideal iBDD'' for a genie-aided iBDD where miscorrections are disregarded
	i.e., the decoder outputs $\lrr$ in case of miscorrection.    

	The iBDD-SR decoding algorithm for product-like codes proposed in \cite{She18,She19Tcom} has similar complexity as that of iBDD. In the following we briefly review iBDD-SR. 
	We consider the decoding of the $i$-th row code. First, BDD is performed and then, in order to
	combine the BDD output with the channel LLRs, the decoded bits are
	mapped according to $0 \to +1$ and $1 \to -1$ if BDD is successful and
	mapped to $0$ in case of decoding failure. The result of this mapping is denoted by $\bar{\mu}_{i,j}^{\mathsf r, (\ell)} \in \{\pm1, 0 \}$ for   bit $c_{i,j}$. 
	The decoder computes $\psi_{i,j}^{\mathsf{r},(\ell)}=
	\BB(\w_\ell \cdot \bar{\mu}_{i,j}^{\mathsf r, (\ell)} + L_{i,j})$
	as the message that is passed to the $j$-th column code. 
	After decoding of all row codes, the same decoding rule is applied to the column codes. (see \cite{She18,She19Tcom} for details).
	\vspace{-1ex}
	\section{Density Evolution for PC Design}
	We use DE to optimize the code parameters in order to optimize the code threshold, i.e., the pre-FEC BER where the code performance curve bends to the so-called waterfall region and reaches  a target post-FEC BER.
	PCs are contained in the ensemble of generalized low-density parity-check (GLDPC) codes as a class of codes on graphs. Thus, the DE for GLDPC ensembles can be used for finding the  threshold of a PC \cite{JianPfister2017}. DE computes the threshold of a GLDPC ensemble by tracking the outbound error probability of the component decoders over iterations of the iterative decoding. The DE for iBDD-SR for GLDPC ensembles is derived in \cite{She19Tcom}. It provides the optimal scaling factors $\w_\ell$ (optimal in an asymptotic sense, i.e., for large block length) and relies on the weight enumerator (WE) of the component code. To achieve a given OH, a common approach is to shorten the component code. For non-shortened BCH codes, the WE is known. Unfortunately, the WE of shortened BCH codes is not know in general, which  prohibits the direct use of the derived DE in \cite{She19Tcom} for code optimization.
	
	
	For iBDD, in the asymptotically large block length regime the average number of errors in the non-shortened BCH component code converges to a limit  \cite{JianPfister2017}. Thus, one can approximate the outbound error probability of the shortened BCH code based on its non-shortened counterpart. The intuition is as follows. Let $p_\text{in}$ and $p_\text{out}$ be the pre-FEC BER and corresponding post-FEC BER of a non-shortened BCH component code $(2^v-1,2^v-vt-1,t)$. Also, let $p^\text{s}_\text{in}$ and $p^\text{s}_\text{out}$ denote pre-FEC and corresponding post-FEC BER of a shortened BCH code with parameters $(2^v-s-1,2^v-vt-s-1,t)$. The non-shortened BCH code is capable of correcting an average portion of $(2^v-1)p_\text{in}$  input errors, yielding to $(2^v-1)p_\text{out}$ output errors. Shortened component codes are usually decoded based on the corresponding non-shortened component decoder, where some of the code bits are known and hence are error-free. Therefore, the same performance is expected for a shortened BCH component code if the average number of input and output errors is the same as that of the non-shortened BCH component code, i.e., 
	$p^\text{s}_\text{in}=\frac{(2^v-1)}{2^v-1-s}p_\text{in}$ and $p^\text{s}_\text{out}=\frac{(2^v-1)}{2^v-1-s}p_\text{out}$. 
	
	For iBDD-SR, we use the same intuition in order to track the outbound error probability of the BCH component decoders and hence computing the threshold of PCs. 
	\section{Numerical and Simulation Results}
	\begin{figure}[t] \centering 
		\includegraphics[width=\columnwidth]{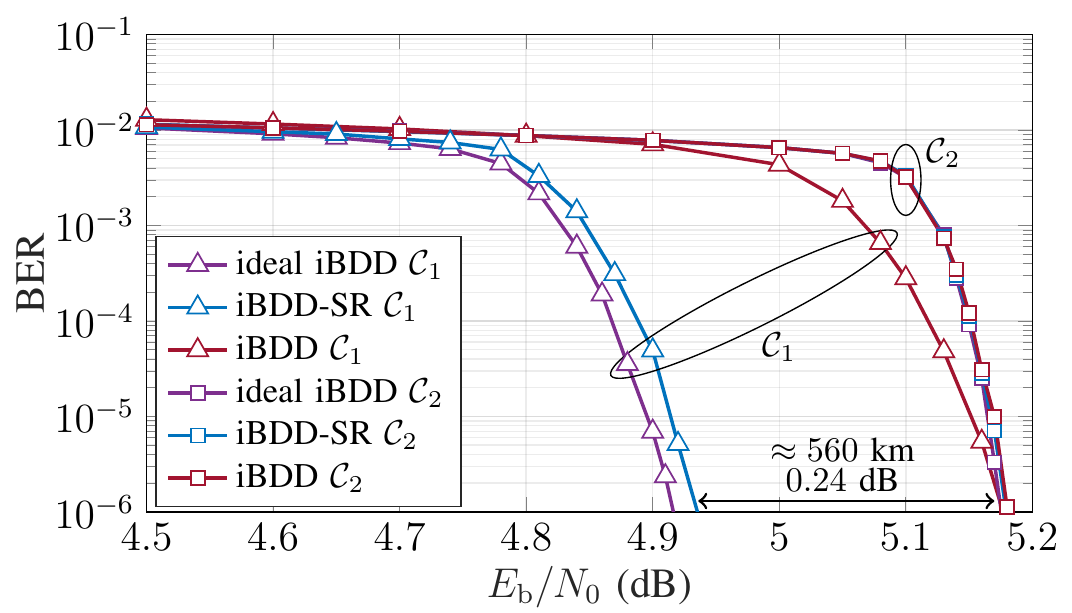}  
		\vspace{-5ex}
		\caption{Performance of PC using iBDD, ideal iBDD, and iBDD-SR for OH of $11.11 \%$ and components $\mathcal{C}_1$ and $\mathcal{C}_2$.}  \vspace{-2ex}
		\label{SimPC1111} 
	\end{figure}
	
	We consider PCs with BCH component codes constructed over GF($2^v$) with $v \in \{8,9,10,11,12\}$ and $t \in \{3,4\}$ as the search space for optimizing the code parameters and different OHs, where $\text{OH} \in \{1/i, i:3,4,\dots,16\}$ \cite{Zhang2014st}. This is a practical search space as for $t>4$ the complexity of the BCH decoder heavily increases and usually codes with $t<3$ suffer from an error floor \cite{Fougstedt2019}. We optimize the code parameters for both iBDD-SR and iBDD to have the best DE threshold for the target post-FEC BER of $10^{-15}$. The optimized parameters are summarized in Table~\ref{tab:Table1}. As can be seen, no code with $v \in \{11,12\}$ gives the best DE threshold. For half of the OHs, the best DE threshold for iBDD requires $t=4$ and for the others iBDD requires $t=3$, whereas $t=3$ gives the best threshold for iBDD-SR and all OHs. 

	In Fig.~\ref{SimPC1111}, we simulate the performance of iBDD, ideal iBDD, and iBDD-SR for a PC with component code of parameters $(9,3,0)$, denoted by $\mathcal{C}_1$, which is the optimized code for $11.11 \%$ OH for both iBDD and iBDD-SR. For iBDD we used $12$ iterations, while for iBDD-SR we used $10$ iterations with $2$ additional iterations of iBDD to correct the errors with high reliabilities \cite{She19Tcom}. As can be seen, iBDD-SR outperforms iBDD by \AS{$0.24$ dB} at a BER of $10^{-6}$. Using the GN model for WDM system with $81$ channels, $32$ Gbaud, $80$ km spanlength, and typical fiber parameters \cite[Table I]{sheikhAIR17}, this gain translates into $560$ km of reach enhancement with QPSK modulation where the original reach is $9680$ km. The error floor due to stoping set of minimal size \cite{Hag18} for $\mathcal{C}_1$ is computed as $10^{-18}$, which confirms that the optimized code can be used for fiber-optic systems. 
	To confirm the poor performance of codes with $v \in \{11,12\}$, we also simulate the performance of the PC with component code $(11,4,1190)$ denoted as $\mathcal{C}_2$ in the figure. One can see that the performance of all algorithms degrades significantly and they perform almost the same. 
	\begin{figure}[t] \centering 
		\includegraphics[width=\columnwidth]{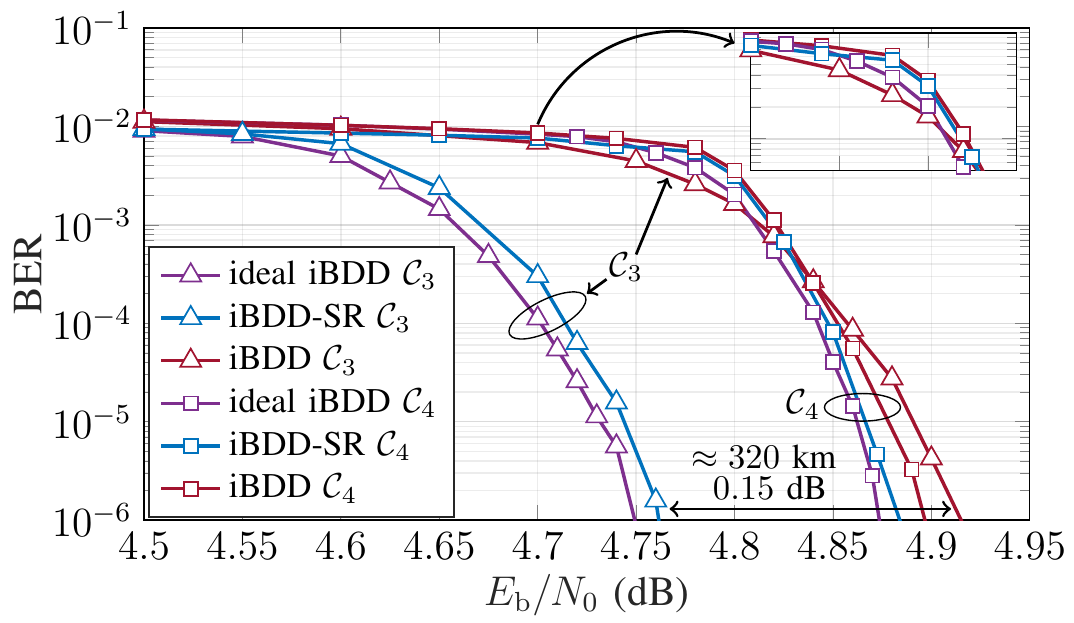}  
		\vspace{-4ex}
		\caption{Performance of PC using iBDD, ideal iBDD, and iBDD-SR for OH of $14.29 \%$ and components $\mathcal{C}_3$ and $\mathcal{C}_4$.}  \vspace{-2ex}
		\label{SimPC1425} 
	\end{figure}  
	
	In Fig.~\ref{SimPC1425}, we simulate the performance of iBDD, ideal iBDD, and iBDD-SR for a PC with component code with parameters $(9,3,93)$, denoted by $\mathcal{C}_3$, which is the optimized code for $14.29 \%$ OH for both iBDD and iBDD-SR. For the sake of comparison, we also simulate the PC with component code $(10,4,404)$ denoted as $\mathcal{C}_4$ in the figure, that has a similar threshold compared to that of the optimized one. One can see that iBDD-SR for both $\mathcal{C}_3$ and $\mathcal{C}_4$ approaches the performance of ideal iBDD. Moreover, the performance improvement of $\mathcal{C}_3$ with iBDD-SR over iBDD yields $320$ km optical reach increase over $10560$ km original reach in the considered WDM system with QPSK.
	Note that $\mathcal{C}_3$ with $n_\mathsf{c}=174724$ is smaller than $\mathcal{C}_4$ with $n_\mathsf{c}=383161$ and also requires $t=3$. We remark that there is an efficient implementation for BCH codes with $t=3$ compared to general BCH codes \cite{staircase_frank}, and also a shorter code requires less circuit area \cite{Fougstedt2019}, hence, interestingly $\mathcal{C}_3$ with iBDD-SR offers better performance with even less complexity compared to $\mathcal{C}_4$ with iBDD. This highlights the importance of code optimization for a given OH. We also remark that although $\mathcal{C}_3$ has the best DE threshold for iBDD and clearly the best performance in the waterfall region (see zoomed part in Fig.~\ref{SimPC1425}), $\mathcal{C}_4$ provides better BER slope and thus gives small improvement at a BER of $10^{-6}$. 
	
\begin{figure}[t] \centering 
		\includegraphics[width=\columnwidth]{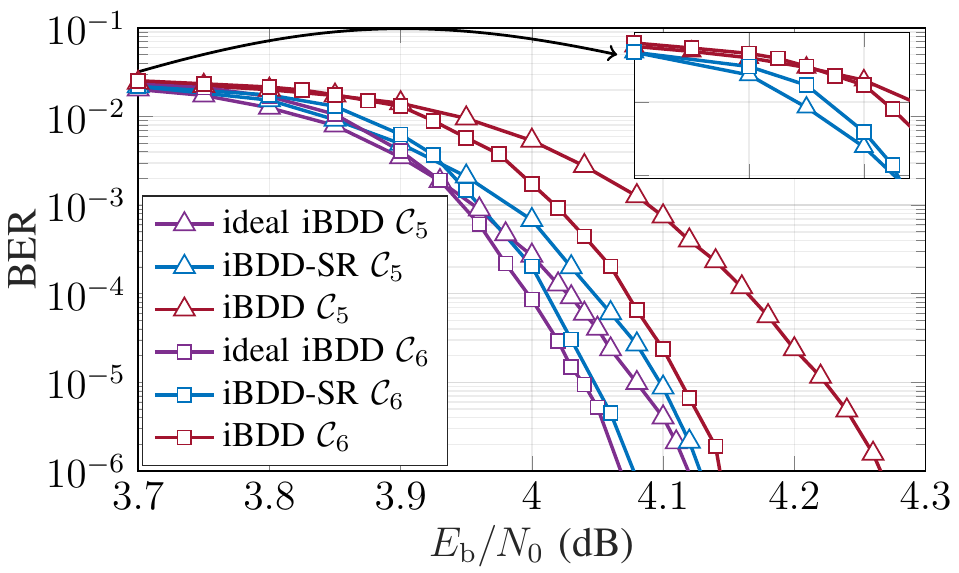}  
		\vspace{-5ex}
		\caption{Performance of PC using iBDD, ideal iBDD, and iBDD-SR for OH of $33.33 \%$ and components $\mathcal{C}_5$ and $\mathcal{C}_6$.}  \vspace{-2ex}
		\label{SimPC3333} 
	\end{figure}

	In Fig.~\ref{SimPC3333}, we simulate the performance of two PCs, named $\mathcal{C}_5$ and $\mathcal{C}_6$, with component codes of parameters $(8,3,76)$ and $(8,4,16)$, respectively. $\mathcal{C}_5$ is the optimized code for iBDD-SR and $\mathcal{C}_6$ for iBDD for $\text{OH}=33.33 \%$.  As can be seen, $\mathcal{C}_5$ provides better performance in the waterfall region than $\mathcal{C}_6$ for iBDD-SR while $\mathcal{C}_6$ gives better better performance in the waterfall region than $\mathcal{C}_5$ for iBDD (see zoomed part in Fig.~\ref{SimPC3333}). At a BER of $10^{-6}$, $\mathcal{C}_6$ gives better performance in both cases due to a larger BER slope. The performance improvement of $\mathcal{C}_6$ with iBDD-SR over iBDD gives $160$ km reach increase over $14720$ km original reach in the considered WDM system with QPSK. We remark that the performance improvement of iBDD-SR with $\mathcal{C}_6$ over iBDD-SR with $\mathcal{C}_5$ is limited \AS{(less than $0.05$ dB)} and is at the cost of significant increase in implementation complexity due to using $t=4$ rather than $t=3$. Therefore, the optimized code $\mathcal{C}_5$ is expected to provide the best performance-complexity trade-off for $33.33 \%$ OH.   
	
	\section{Conclusion}
	We addressed the design of binary PCs under iBDD-SR for different OHs using DE to optimize the performance in the waterfall region, as corroborated by simulation. We showed that the performance of the optimized codes with iBDD-SR approaches that of the genie-aided miscorrection-free decoder. Also, we showed that the optimized PCs with iBDD-SR yield an optical reach enhancement up to $5.8 \%$ of the original reach achieved by iBDD for a WDM system with QPSK modulation. The DE as well as the simulation results confirmed that the parameters of component codes have a prominent effect on the performance of binary PCs. Overall, we conclude that for all OHs, binary PCs with iBDD-SR and
	$t = 3$ provides the best performance-complexity
	trade-off. 

	\vspace{-1ex}
	\section*{Acknowledgment}
	
	\scriptsize{The work of A. Sheikh and A. Alvarado has received funding from the European Research Council (ERC) under the European Union's Horizon 2020 research and innovation programme (grant agreement No 757791). The work of A. Sheikh is partially supported by the Knut and Alice Wallenberg Foundation.}
	
	\section*{References}
	\normalsize

\end{document}